\documentclass[a4paper,11pt]{article}

\usepackage{contribution}

\newcommand{\weblink}[2][]{%
    \ifthenelse{\equal{#1}{}}%
    {\textnormal{\url{#2}}}%
    {\textnormal{\href{#2}{#1}}}%
}

\newcommand{\acknowledgements}[1]{%
  \bigskip\bigskip
  \textsf{\textbf{\Large Acknowledgements}} \\[2ex]
  {#1}
  \bigskip
}

\def\beq{\begin{equation}}
\def\eeq#1{\label{#1}\end{equation}}
\def\eeqn{\end{equation}}

\def\beqa{\begin{eqnarray}}
\def\eeqa#1{\label{#1}\end{eqnarray}}
\def\eeqan{\end{eqnarray}}

\let\bar=\overbar

\def\Dslash{\not{\hbox{\kern-4pt $D$}}}
\def\dslash{\not{\hbox{\kern-2pt $\del$}}}

\def\msb{{\bar{\ssstyle M \kern -1pt S}}}

\newcommand{\contribution}[7][]{%
  \clearpage
  \thispagestyle{plain}
  \ifthenelse{\equal{#1}{}}
  {\hypersetup{pdftitle={#2}}}
  {\hypersetup{pdftitle={#1}}}
  \hypersetup{pdfauthor={{#3} {#4}}}
  {\centering\normalfont\LARGE\bfseries\sffamily #2 \par\nobreak}
  \lhead{}
  \chead{%
    \textit{\footnotesize XIV International Conference on Hadron Spectroscopy
      (\weblink[\textit{hadron2011}]{http://www.hadron2011.de}), 13-17 June 2011, Munich, Germany}%
  }
  \rhead{}
  \bigskip
  \begin{center}
    {#3} {#4}\ifthenelse{\equal{#6}{}}{}{\footnote{\weblink[#6]{mailto:#6}}}
    \ifthenelse{\equal{#7}{}}{}{#7} \\
    \textit{#5}
  \end{center}
  \bigskip
}

\renewcommand{\abstract}[1]{%
  \begin{center}
    \begin{minipage}{0.85\textwidth}
      \begin{footnotesize}
        #1
      \end{footnotesize}
    \end{minipage}
  \end{center}
  \bigskip
}

\begin{document}

%
%
%
%
%
{  


%

\contribution[Partial Wave Analysis using GPUs]  
{Partial Wave Analysis using Graphics Cards}  
{Niklaus}{Berger}  
{Institute of High Energy Physics \\
  Chinese Academy of Sciences \\
  19B Yuquan Lu\\
  100049 Beijing, CHINA}  
{nberger@ihep.ac.cn; now at Physics Institute, University of Heidelberg}  
{}
%

\abstract{%
  Partial wave analysis is a key technique in hadron spectroscopy. The use of unbinned likelihood fits on large statistics data samples and ever more complex physics models makes this analysis technique computationally very expensive. Parallel computing techniques, in particular the use of graphics processing units, are a powerful means to speed up analyses; in the contexts of the BES~III, Compass and GlueX experiments, parallel analysis frameworks have been created. They provide both fits that are faster by more than two orders of magnitude than legacy code and environments to quickly program and run an analysis. This in turn allows the physicists to focus on the many difficult open problems pertaining to partial wave analysis. 
}
%

\section{Introduction}

The advent of precise lattice calculations of hadron spectra (e.g.~\cite{Dudek:2011tt}) together with very high statistics data samples from experiments like BES~III and Compass and, in the near future, GlueX and Panda, opens inroads to a deeper understanding of the bound states and resonances of quantum chromodynamics (QCD). Extracting resonance properties from experimental data is however far from straightforward; resonances tend to be broad and plentiful, leading to intricate interference patterns. In such an environment, simple fitting of mass spectra is usually not sufficient and a \emph{partial wave analysis} (PWA) is required to extract resonance properties. In such an analysis, the full kinematic information is used and fitted to a model of the amplitude in a partial wave decomposition. In the cases discussed here, the model parameters are determined by an unbinned likelihood fit to the data. As models become ever more involved and huge data samples are (becoming) available, these likelihood fits are computationally very expensive. As it is exponentially hard to determine whether the fit (usually with a large number of free parameters) has found the searched for global minimum of the likelihood or one (of usually many) local minima and as there tends to be a lot of freedom in the model, many fits are required to obtain a certain degree of certainty on the stability and adequacy of the result. As most of these fits are performed sequentially, the turnaround time for a single fit, together with the effort required to code the amplitude model, determines the total analysis time.

In the context of the BES~III, Compass and GlueX experiments, software frameworks for partial wave analysis addressing both the amplitude code and speed issues have been devised. Large speed-ups were obtained by parallelizing computations, particularly by the use of \emph{graphics processing units} (GPUs). At the same time, the use of a common software base allows for a quick creation of analysis code for new channels. This in turn allows the physicists to focus on the difficult open problems, both of technical (large number of free parameters, correlations, detector resolution etc.) and physical (adequacy of models) nature.

\section{Partial Wave Analysis as a computational problem}

In a typical PWA (we use the simple radiative decay $J/\psi \longrightarrow \gamma X, X \longrightarrow K^+K^-$ \cite{Bai:2003ww} as an example), the decay is modelled by coherently summing the contributions from a variety of intermediate resonances $X$. The relative magnitudes and phases of these resonances are determined from a fit and the fit result is compared to the data. The set of resonances and their properties are changed until a sufficient agreement with the data is found.

The intensity $I$ (relative number of events) at a particular point $\Omega$ in phase space can be expressed as
\begin{equation}
	I(\Omega) = \left|\sum_{\alpha} V_{\alpha}A_{\alpha}(\Omega)  \right|^{2},
\end{equation}
where the sum runs over all intermediate resonances $\alpha$, $V_{\alpha}$ is the complex production amplitude for $\alpha$ (and the main free parameter in the fit) and $A_{\alpha}(\Omega)$ the complex decay amplitude at a particular point in phase space. The likelihood for a particular model is
\begin{equation}
	\mathcal{L} \propto \prod_{i = 1}^{N_{Data}} \frac{I(\Omega_{i})}{\int \epsilon(\Omega) I(\Omega) d\Omega},
\end{equation}
where the product runs over all $N_{Data}$ events in the sample and the integral is proportional to the total cross section, corrected for the detector efficiency $\epsilon(\Omega)$. The integral is usually performed numerically by summing over a large number $N_{MC}^{Gen}$ of simulated events (Monte Carlo, MC) generated evenly in phase space\footnote{In case of a very uneven phase-space population, e.g.~due to the presence of narrow resonances, it is advisable to generate a non-equidistributed MC sample and weight events to a flat distribution in order to reduce the sampling error caused by the strongly varying amplitude}. The limited acceptance and efficiency of the detector can be taken into account by summing only over simulated events that pass the reconstruction and analysis cuts. In a fit, a maximum of the logarithm of the likelihood, corresponding to the best set of parameters for the used model is searched for\footnote{As most fit programs search for a minimum as opposed to a maximum, $-\ln{\mathcal{L}}$ is fed to the minimizer.};
\begin{equation}
	\ln{\mathcal{L}} \propto \sum_{i =1}^{N_{Data}} \ln\left( \sum_{\alpha} \sum_{\alpha\prime} V_{\alpha}V_{\alpha\prime}^{*} A_{\alpha}(\Omega_{i}) A_{\alpha\prime}^{*}(\Omega_{i})\right) \nonumber 	-  \sum_{\alpha} \sum_{\alpha\prime}  V_{\alpha}V_{\alpha\prime}^{*} \left( \frac{1}{N_{MC}^{Gen}} \sum_{j=1}^{N_{MC}^{Acc}} A_{\alpha}(\Omega_{j}) A_{\alpha\prime}^{*}(\Omega_{j})\right).
\end{equation}
The first sum runs over all data events, the second over all MC events. If the widths and masses of resonances are kept constant in the fit (i.e.~the $V_{\alpha}$'s are the only free parameters), the last (inner) bracket and the $A_{\alpha}(\Omega_{i}) A_{\alpha\prime}^{*}(\Omega_{i})$ term for each data event can be pre-calculated.

The number of floating point operations required is dominated by the sum over the data events and scales with $N_{iterations} \times N_{data} \times N_{waves}^2$, whilst the lookup table takes up storage space scaling with $N_{data} \times N_{waves}^2$. 
The storage space problem can be addressed by increasing the memory of the relevant machine ($\approx$1.5~GB are required per million events for a model with 20 partial waves), or with appropriate caching and staging mechanisms for data samples with several million events. If the required floating point operations are performed sequentially, the time required can however become prohibitively long.

Floating point precision can be an issue --- the minimizing programs tend to fail if the delivered precision does not correspond to the expected one. For large amounts of data and simulated events however, the precision of the individual amplitude is not decisive, it can safely be computed in single precision. The final sum however has to be computed in double precision and the numerical result is strongly dependent on the summing algorithm; tree sums as usually implemented on GPUs perform much better here than accumulating loop summing employed in traditional CPU based programs. 

There is no way of knowing whether the fit found just a local or the searched for global maximum of the likelihood. To gain confidence in the result, the fits are usually repeated with various sets of starting parameters. In addition, various models have to be tried out, especially in the study of possible new resonances and systematic effects. The thousands of fits needed in a typical partial wave analysis should thus be as fast as possible, especially as feedback from the physicist is required between the fits and they thus have to be run in sequence. 

\section{Parallel partial wave analysis}

Most computing problems in particle physics are trivially parallel, as data consists of independent events. The traditional approach to parallelism is thus to treat different subsets of events on different cores/machines/in different locations and then have a relatively lightweight piece of code perform a synthesis. For PWA, this synthesis (namely the likelihood sum) has to be performed very frequently (for every fit iteration) and thus network latency quickly dominates for distributed architectures. As the calculations per event are relatively simple and exactly the same for all events, it is desirable to have a very large number of (simple) cores available in a single machine. This kind of architecture is provided by graphics processing units (GPUs). These devices were originally developed for the use in 3D games, calculating realistic colour shades of pixels. The large and performance-hungry game market has produced single chips with 1600 individual floating point units, which are available at prices of a few hundred Euros.

The potential of GPUs for scientific computing was quickly discovered, early efforts were however hampered by the lack of a suitable interface and most work was done via the \emph{OpenGL} graphics interface. The two major GPU vendors, Nvidia and ATI then started their own GPU computing frameworks, named \emph{CUDA} \cite{CUDA} and \emph{ATI Stream} \cite{Buck2004, AMDBrook, AMDStreamOverview} respectively\footnote{The ATI stream framework encompassed at various times different interfaces with different levels of abstraction and is nowadays also used for the ATI \emph{Open CL} implementation.}. In recent years, a platform independent standard called \emph{OpenCL} \cite{OpenCL,OpenCL1} has emerged. The successive choices of GPU interface for the BES III PWA framework and the final transition to \emph{OpenCL} are described in \cite{Taipeh}.

In 2007 we started tests of GPUs for partial wave analysis at the BES~III experiment; it quickly turned out that their architecture was very well suited to the task at hand --- in fact, custom PWA hardware would very much look like a GPU, minus the display port --- and that speed-ups with regards to a reference FORTRAN implementation of more than two orders of magnitude were obtainable \cite{Berger:2010zz}. The framework now provides facilities for amplitude calculation, minimization and plotting and is widely used for analyses at BES~III. It continues to be developed and is available at \cite{GPUPWA_sourceforge}.

The group at Indiana University at around the same time started creating a framework for PWA on multi-core machines and clusters, using \emph{OpenMP} for inter-process communication. This framework, aimed at the CLEO-c, BES~III and GlueX experiments shows beautiful scaling behaviour and now also harnesses GPUs using \emph{CUDA}. It is available at \cite{AMPTOOLS_sourceforge}.

Yet another framework, tightly integrated with the \emph{root} data analysis toolset \cite{Brun1997}, was developed in Munich for use in the COMPASS experiment and now also incorporates GPU assisted fitting. It is available at \cite{ROOTPWA}.

A more ambitious effort for a very general framework for the PANDA experiment was started in Bochum and is still in its early stages, is however also aimed at an eventual parallel implementation.

The multitude of available frameworks demonstrate the great interest in the field, allow for cross checks and show that speed is indeed an important issue in PWA, but one that can be tackled with an appropriate combination of hard- and software.

\section{Open technical problems in partial wave analysis}

Probably the most difficult problem in PWA is the appropriate modelling of the physics processes, taking into account the unitarity of the $S$-matrix. There is currently a lot of theoretical work done in this area (see e.g. \cite{trento}); a completely satisfying treatment has however only been achieved for low energy $\pi\pi$ scattering. There is in addition a large range of technical/experimental problems as yet unsolved; they shall be discussed in the following.

\paragraph{Resolution}
Detectors have a finite resolution and the reconstructed kinematics of an event are thus different from the true ones. The unbinned approach to PWA precludes the use of unfolding techniques. For resonances with a large width, this is not particularly problematic; in cases however where resolution and width are of comparable size (a good example for most detectors is the $\phi$ meson), it is impossible to describe interference with other resonances and the reconstructed mass distribution with the same functional form. The only (and rather unsatisfactory) remedy is excluding the mass regions around narrow resonances from the fit, in the hope that the effects in the tails are not to severe.

\paragraph{Goodness of fit}
For unbinned distributions in a multi-dimensional variable space, there are is no effective goodness of fit test such as a $\chi^2$ test; of course it is possible to perform tests on all sorts of binned projections --- if these look bad, the fit is indeed bad, but unfortunately the opposite is not necessarily true.

\paragraph{Choice of waveset}
The choice of intermediate resonances is somewhat arbitrary; not dissimilar to the goodness of fit issue it is easy to tell if the waveset is insufficient, but a priori not knowable whether it is complete.

\paragraph{Fits with large numbers of parameters}
PWA fits have at least three (more often four, in more advanced models also more) free parameters per resonance, leading to fits with a very large number of parameters. In such fits it is impossible to know whether a minimum identified by the minimizer is just a local minimum or the searched for global one. Many heuristic methods (such as repeating fits with randomly chosen initial values) have been developed to gain confidence in a particular minimum; they however become excessively time consuming for large parameter spaces and will never provide certainty. 

\paragraph{Fits with complex parameters}
PWA fits usually have complex parameters. As minimizers only deal with real numbers, some representation has to be chosen; however, both the Cartesian and the polar representations have issues. The Cartesian representation has the disadvantage of being disconnected from the physical parameters of magnitude and phase and leads to large correlations (and thus bad fit convergence) in cases where the phase of a resonance is weakly constrained. The polar representation avoids these problems at the cost of introducing a lower boundary on the magnitude and a periodic phase parameter - both of which have their separate issues in the minimizers. A proper treatment of complex numbers and their derivatives inside the minimizers would be extremely helpful.

\paragraph{More control over the fitters}
The standard interfaces to fitters like \emph{Minuit} \cite{Minuit} and \emph{Fumili} \cite{Fumili} give the user relatively few control over the amount of additional checks and error calculations performed by the minimizers. In many cases however, most of the computing time is spent on just these. Options to perform only what is really needed at the current state of analysis would be very helpful.

\paragraph{Less control over the GPUs}
The two most widely used GPU frameworks today, \emph{CUDA} and \emph{OpenCL}, both have a fairly low level interface, giving the user a lot of control over how exactly the parallelism is to be implemented --- for most particle physics applications it is however completely irrelevant which events are processed simultaneously, as long as all of them are processed fast. This could be achieved through a more comfortable high level interface.

\paragraph{Impact of results}
The last issue is sociological rather than technical in nature, but tightly linked with many of the technical issues. In very generalized terms, ``standard'' results, where PWA is used to determine quantum numbers of a bump visible in the mass spectrum and these turn out to be in accordance with expectations are easily accepted by the community (more or less independently of the sophistication of the analysis), ``surprising'' results, where resonances not directly visible in the mass spectrum are exposed by PWA and turn out to have unexpected (exotic) quantum numbers are usually not believed (again more or less independently of the sophistication of the analysis). As long as the issues listed above remain, PWA has a very hard time to provide the required \emph{extraordinary evidence for extraordinary claims}.

\section{Conclusions}
\label{sec:Conclusions}

Partial wave analysis is a key tool in hadron spectroscopy, of great importance to current and future experiments such as BES~III, Compass, GlueX and Panda. PWA is computationally expensive and thus potentially slow --- this can be overcome by applying parallel programming techniques and running the analyses on massively parallel devices such as GPUs. The success in this area gives us both the tools and the hope that we can solve or mitigate many of the other technical problems facing PWA, thus laying a foundation for a new golden age of hadron spectroscopy.

\acknowledgements{%
  I am grateful to Matthew Shepherd, Ryan Mitchell and Sebastian Neubert for many insightful discussions about PWA fitting, hadron spectroscopy and fast code. The many users of the GPUPWA code in BES~III have helped to move the project in the right direction by providing feedback, feature requests and bug reports. 
  
This work has been financed by the Chinese Academy of Sciences and the Swiss National Science Foundation.
}


%

}  


\end{document}